# Efficient and Accurate In-Database Machine Learning with SQL Code Generation in Python

Michael Kaufmann
*Departement Informatik*
*Hochschule Luzern)*
Rotkreuz, Switzerland
m.kaufmann@hslu.ch

Gabriel Stechschulte
*Departement Wirtschaft*
*Hochschule Luzern*
Luzern, Switzerland
gabriel.stechschulte@stud.hslu.ch

Anna Huber
*Departement Informatik*
*Hochschule Luzern*
Rotkreuz, Switzerland
annamagdalena.huber@hslu.ch

*Abstract*— Following an analysis of the advantages of SQL-based Machine Learning (ML) and a short literature survey of the field, we describe a novel method for In-Database Machine Learning (IDBML). We contribute a process for SQL-code generation in Python using template macros in Jinja2 as well as the prototype implementation of the process. We describe our implementation of the process to compute multidimensional histogram (MDH) probability estimation in SQL. For this, we contribute and implement a novel discretization method called equal quantized rank binning (EQRB) and equal-width binning (EWB). Based on this, we provide data gathered in a benchmarking experiment for the quantitative empirical evaluation of our method and system using the Covertype dataset. We measured accuracy and computation time and compared it to Scikit Learn state of the art classification algorithms. Using EWB, our multidimensional probability estimation was the fastest of all tested algorithms, while being only 1-2% less accurate than the best state of the art methods found (decision trees and random forests). Our method was significantly more accurate than Naive Bayes, which assumes independent one-dimensional probabilities and/or densities. Also, our method was significantly more accurate and faster than logistic regression. This motivates for further research in accuracy improvement and in IDBML with SQL code generation for big data and larger-than-memory datasets.

## I. INTRODUCTION

Today, most mobile phones have more storage capacity than the volume of data that common machine learning (ML) algorithms can process. There are a variety of existing methods for ML and data mining, which are available in languages like R, Python or Java and in environments like CRAN, PyPi and Weka. However, most of these methods and tools only work for data sets that can be processed in-memory. The problems of big data in the ML area begins with volumes of a few gigabytes.

There are several approaches to circumvent the memory bottleneck. Incremental algorithms build models record by record, parallelization techniques deploy several hardware computers at the same time, and in-memory methods use massive amounts of RAM, for example with GPU computing. There is also the possibility to extend the RAM to disk, for example with the package ff for R. In this work, however, a database-based, and specifically SQL-based, approach to big data shall be investigated.

The Seattle Report on Database Research published 2020 [1], which was written by database researchers including Michael Stonebraker (MIT) and Phil Bernstein (Microsoft Research), describes in-database machine learning (IDBML) as an important research direction for database research in the coming years. IDBML has many advantages, because the calculation of ML models is done directly on the database server, without data migration. This offers advantages regarding data protection and data volume.

IDBML can be accomplished using SQL code generation. If a ML approach can be completely or partially implemented in SQL, this has several advantages that are outlined below.

- Code-To-Data: The data is processed where it already exists and does not need to be migrated.
- Data Ethics: Data never leaves the SQL DB server during processing. This has enormous advantages in terms of data protection and privacy.
- Scalability: The implementation via a standardized SQL interface offers the possibility of connection via Hive or Drill, so that it works on parallelizable, massively scalable NoSQL databases.
- Reusability: A database management system (DBMS) can natively process amounts of data that go far beyond main memory. This functionality can be reused in ML algorithm implementation.
- Scope: SQL is an international standard supported by hundreds of database types in Millions, possibly Billions of information systems.

In this sense, in the research described here, a specific ML procedure, namely multidimensional histogram (MDH) probability estimation, was implemented using SQL code generation, and made interactively available via Python. Also, we examined our ML method and system for its properties in comparison to existing procedures.

In the following sections, we describe the state of technology in IDBML with SQL in Section II; the math behind our approach in Section III; the architecture and process flow of our software implementation in Section IV; The lab experiments and their results in Section V; a discussion and interpretation of results in Section VI; our conclusions supported by data in Section VII, and points of further research in Section VIII.

## II. STATE OF TECHNOLOGY

IDBML is an important research direction for not only database research, but also ML in general. A detailed discussion of the role of databases in machine learning can be found in the overview by Kumar et al. (2017)[2], however, they do not discuss the relationship between SQL and IDBML. In the following, a few existing solutions of combining IDBML with SQL are summarized.

*A. SQL Extensions for IDBML*

Well-known commercial suppliers of database technology have contributed to the in-database ML realm with products and services. Microsoft SQL Server ML Services lets users run Python and R code from within the SQL server. Oracle ML Notebooks and Google BigQuery have extended the SQL language to implement machine learning in queries, each with their own proprietary dialect. Amazon Aurora Machine Learning only creates a syntactical link to external ML tools such as Amazon Comprehend or SAGEmaker. Teradata SQL Engine [3] allows the interactive use of Python ML libraries directly inside the SQL database without data migration. However, all these products are proprietary, and the syntax is specific for each database brand. Likewise, open-source projects have developed such as MADlib [4] to address the fundamental problem of big data scalability with open-source methods and tools. MADlib implements ML algorithms using user defined functions for PostgreSQL and Greenplum. Again, even though the implementations are open source, the syntax is specific for PostgreSQL, and MADlib does not work on other systems supporting the SQL standard interface. Other open source and research prototypes exist such as SQLflow [5] which extends the SQL language to support ML workflows between database systems and ML libraries. It builds upon the descriptive paradigm, describing what the outcome of the computation should be, instead of stating how the computation is to be achieved. Similarly, Bordaweka has integrated Word2Vec vector space experimentally into an existing standard SQL query processing system and exposed the vector-based operations via SQL queries. [6].

*B. Implementation of ML in SQL*

Another approach, though less researched and with less open-source contributions, is implementing ML algorithms directly in SQL using SQL syntax. SQL is a query language designed for querying and analyzing data, while Python or R are general-purpose languages that promote data analysis and experimentation. Given the objective of the SQL query language, implementation of more complex ML algorithms via SQL syntax may be feasible, and may look trivial at first sight, but turn out to be complex and difficult, especially with respect to efficiency optimization. Thus, open-source ML algorithms constructed via SQL syntax are not readily available. Yet, there are several research protypes demonstrating the performance gains possible by at least partially computing ML by leveraging the power of server-side SQL query processing.

As early as 1999, Chaudhur, Fayyad and Berhardt from Microsoft Research [7] proposed to implement ML in SQL and demonstrated the feasibility and efficiency with a decision tree implementation. In the year 2000, Ordonez and Cereghini [8] presented three implementations of expectation maximization (EM) clustering in ANSI SQL. They noted that while coding ML algorithms in SQL may look trivial, it turned out to be a challenging problem when handling large datasets with high dimensionality. Also, SQL supports relational algebra, but not linear algebra, because SQL tables are not matrices; a general problem that calls for solutions. In 2004, Bentayeb [9] proposed this approach for IDBML and implemented an intrusion detection system (IDS) decision tree partially using SQL. In the same year, Ordonez [10] demonstrated K-Means clustering in SQL. In 2009, Navas and Ordonez [11] showed how to implement principal component analysis (PCA) using singular vector decomposition (SVD) in SQL. In 2010, Ordonez and Pittchaimalai [12] published their SQL implementation of Bayesian Classifiers. Although their SQL code is four times slower than the corresponding C++ program, SQL scales linearly and in contrast to C++, it can compute ML on datasets much larger than the main memory. In 2012, Nagarjuna and Reddy [13] implemented Bayesian classifiers in SQL and proposed to generate SQL code from a host language, thus achieving Turing-completeness together with SQL larger-than-memory computations, a "best-of-both-worlds" approach we will build upon in our research using Python to generate SQL code. Moreover, Bayesian classifiers are particularly suited for SQL code generation because SQL aggregation and grouping functionalities can efficiently compute multidimensional contingency tables, the basis for every Bayesian statistic. In 2017, Marten and Heuer [14] have implemented Hidden Markov Models in SQL. In 2019, Marten and Heuer [15] implemented Fourier transforms in SQL. In 2020, Julia Glick from Mode Analytics, inc., has published a white paper [16] describing how to do least squares regression using general SQL queries. This method can be automated using Python and Jinja2 templates – a task that a bachelor's student supervised by one of the authors is currently tackling.

### III. FORMAL METHODOLOGY

MADlib rests on three skills called "MAD" skills: Magnetic, agile and deep data analysis [17]. However, the software library is implemented using a DBMS-specific SQL dialect and DBMS-specific procedural extensions (pl/pgsql). Because of the reasons stated in the introduction, we are more interested in Standard-SQL implementations of ML algorithms so that they can be run on any system supporting the SQL standard interface. Therefore, instead of MAD skills, we propose *SANE skills* for data analysis in the area of IDBML as an acronym for *SQL Analytics for Numerical Estimation*. The idea of numerical estimation is to solve a mathematical problem in a practical, goal-oriented way that is as efficient as possible, without the claim of being theoretically correct or optimal. Accordingly, the math for our approach is rather trivial, but can be implemented in a scalable way using SQL code generation.

In supervised learning, a training data set $D$ is a $m \times (n+1)$ table with $n$ columns $X_1, \ldots, X_n$ indicating the characteristics of the observations, called attributes, and a column $Y$ indicating the class of the observations, called the target. The rows represent data elements $r_i = (X_{i1}, \ldots, X_{in}, Y_i)$. For a prediction $\hat{Y}_\alpha$ of the class value for a new data element $r_\alpha$, we want to estimate the value of the unobserved target variable that has the highest probability of occurring, based on the observed attribute values:

$$\hat{Y}_\alpha = \arg\max_y P(y|X_{\alpha 1}, \ldots, X_{\alpha n}) \qquad (1)$$

According to Bayes rule, we have

$$\hat{Y}_\alpha = arg\ max_y \frac{P(X_{\alpha 1}, \ldots, X_{\alpha n}, y)}{P(X_{\alpha 1}, \ldots, X_{\alpha n})} \qquad (2)$$

And because $P(X_{\alpha 1}, \ldots, X_{\alpha n})$ is constant for $r_\alpha$,

$$\hat{Y}_\alpha = arg\ max_y\ P(X_{\alpha 1}, ..., X_{\alpha n}, y) \quad (3)$$

Therefore, predicting a class value depends on estimating the joint probabilities of the combination of all attribute and target values – estimating a *multidimensional probability*. The probability $P$ of an attribute value combination can be estimated by the percentage $P_D$ of the training dataset $D$ that corresponds to this pattern.

$$P_D(x_1, ..., x_n) := \frac{1}{m} \left| \bigcap_{j=1}^{n} \{r_i \in D | X_{ij} = x_j\} \right| \quad (4)$$

For categorical variables, this probability estimation can be achieved simply by counting the number of occurrences per value combination. For numerical variables, there are different possibilities such as density estimation or discretization. With regard to SANE skills described above, we choose a computationally simple variant: a histogram probability estimation method. We discretize the numerical variable into n bins. There are many different variants for discretization, such as equal frequency binning, equal width binning, or complex supervised variants. In our approach, we designed computationally simple discretization methods: First, *equal quantized rank binning (EQRB)*, using following formula 5a,

$$Q(N_{ij}) = ceil(b * rank(N_{ij})/m) \quad (5a)$$

where $Q(N_{ij})$ is the resulting quantized bin index, $N_{ij}$ is a numerical value of attribute $j$ for record $i$, $b$ is a parameter specifying the desired number of bins, $m$ is the number of rows in the dataset, $ceil(x)$ is a function that returns the closest integer larger than $x$ for a floating point value $x$, and $rank(N_{ij})$ is a window function that returns the row number of a value $N_{ij}$ in the ordered list of all values of attribute $N_j$ in the dataset. Second, equal width binning (EWB), using formula 5b,

$$Q(N_{ij}) = ceil\left(b * \frac{N_{ij} - min_D(N_{ij})}{max_D(N_{ij}) - min_D(N_{ij})}\right) \quad (5b)$$

where $Q(N_{ij})$, $N_{ij}$, $b$ and $ceil(x)$ are defined as in Formula 5a, and $min_D(N_j)$ and $max_D(N_j)$ are the minimal and maximal values, respectively, of attribte $j$ in training set $D$. Therefore, in a mixed dataset with $a$ categorical variables $c_j$, $0 \leq j \leq a$ and $b$ numerical variables $n_k, 0 \leq k \leq b$, the probability is estimated according to the following formula:

$$P_D(c_1, ..., c_a, q_1, ..., q_b) :=$$
$$\frac{1}{m} \left| \bigcap_{j=1}^{a} \{r_i \in D | C_{ij} = c_j\} \bigcap_{k=1}^{b} \{r_i \in D | Q(N_{ik}) = q_k\} \right| \quad (6)$$

Now, this MDH probability estimation (PE) works for low dimensional data only, because of the curse of dimensionality. Therefore, the most relevant attributes need to be selected. In our approach, we ranked the attributes regarding their mutual information with the target variable, estimated using the following formula.

$$I(X, Y) = \sum_x \sum_y P_D(x, y) log_2\left(\frac{P_D(x,y)}{P_D(x)P_D(y)}\right) \quad (7)$$

Again, x and y are either categories for categorical attributes, and the probabilities are estimated according to Formula 4; or, dealing with numerical attributes, x and y represent bins calculated according to Formulae (5a) or (5b), and the probabilities are estimated according to Formula (6).

IV. THE SANELIB SOFTWARE PROTOTYPE

We implemented the method described above, first in plain SQL on a MySQL 8 server, and then we generalized and parametrized the SQL for code generation using Python. The use case and software architecture are illustrated in Figure 1. The numbers in the figure correspond to the numbers in the following description.

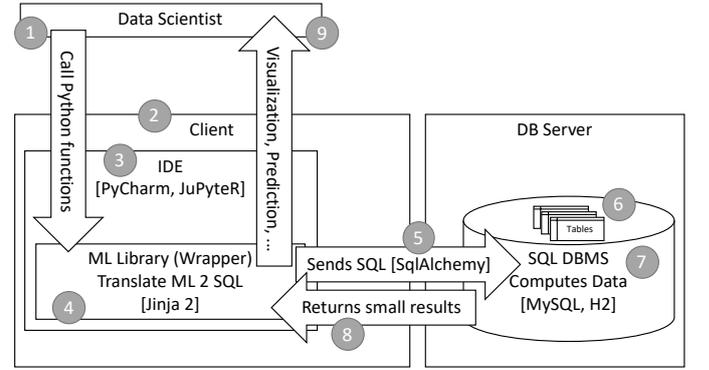

*Figure 1: Software components and use case*

(1) The data scientist expects to do machine learning by calling Python functions with appropriate parameters.
(2) The data scientist then interacts with his client terminal.
(3) On the client, the code is usually run on a Python IDE such as PyCharm or Jupyter Lab.
(4) Our ML library is a wrapper that generates machine learning algorithms in SQL code using Jinja 2 templates.
(5) The ML library sends the generated SQL code to the database server using SQLAlchemy. Thus, the library works on any DBMS that supports a given set of functions of the SQL standard.
(6) The training and test data are stored on the database server that supports a standard SQL interface.
(7) The SQL DBMS computes the heavy load of the calculations for the ML procedure. Leveraging the mature technologies of DBMS such as paging, indexing, and materialized views, the DBMS can compute the ML procedures faster on bigger datasets than the client could do in-memory.
(8) The DB server sends back a small table of aggregated results for further processing. If needed, steps 7 and 8 can be repeated iteratively.
(9) The data scientist can read and utilize the analysis results for prediction and visualization.

The software stores all information for an analysis in several tables with the same prefix called the Model-ID. The computation of the model from the training data is accomplished iteratively using generalized SQL templates, identified by a Template-ID, which are parameterized by the user via Python parameters. This method is used for the training phase and prediction phase as well. For all computation steps, a SQL template is executed. For most computation steps, a materialized view is generated using the following SQL code pattern, as it has turned out to be most efficient procedure:

```
CREATE TABLE <Model-ID>_<Template-ID> AS
<SQL-template-with-parameters-applied>
```

The training method consists of the following steps (with Template-ID in parentheses): (QT) Quantization of the training table; (QMT) Generating of quantization metadata for training table; (M) computing a multidimensional contingency table.

Likewise, the prediction method consists of the following steps: (QE) quantization of metadata for the evaluation table (QE_IX) creating an index on the metadata table; (_P) Generating the prediction table.

Each step will be described in the following paragraphs. The training phase consists of three steps and can be initiated using the `train` subroutine. First, the numeric attributes of training table are quantized. The quantization of the training table template is selecting, according to the user's inputs, the attributes of the target variable, and its associated feature variable(s) and is putting numeric attributes into their respective bins according to Equation (5) above. The quantization of metadata for the training table is generated from the previous template and is determining the local minimum and the local maximum of each respective bin. Likewise, global minimum and maximum values are determined. Lastly, the contingency template models the frequency distribution of the target variable with its respective feature variable(s) based off the previous two tables. The resulting multi-dimensional contingency table counts the number of occurrences for every combination of predictive and (quantized) target attributes occurring in the training data. This model-table acts as the predictive model during the prediction process.

The prediction phase consists of three steps and can be called by the `predict` subroutine. First the numeric attributes of the evaluation table are quantized. The evaluation table provides new instances that have not been used in the training phase, either to evaluate the model or to make actual predictions in real world scenarios. Second, an index on the quantified features of the evaluation table is performed to make the queries faster. Third, a prediction table is generated by joining the model table to the quantized evaluation table, thus being able to evaluate the multidimensional probability of each target class given the feature combination, according to Equation (6) and therefore to infer the most probable class for every instance according to Equation (3). At the end of the prediction phase, the prediction table named `<Model-ID>_P` contains, for each record of the supplied evaluation table, the predicted class value. In case there are new feature combination that have not been observed in the training phase, the majority class is predicted. This predicted value can be used either for calculating model accuracy in experiments, or for decision support in real applications.

Furthermore, our software has a method (`rank`) that can compute the mutual information for every column of the training table with the target attribute. Also, we have implemented a 1d and 2d visualization method (`visual1D`, `visual2D`), resulting in graphs illustrated in Figure 2. For this we have used Plotnine, a port of Ggplot from R to the Python language. The source code of the resulting software package, the SANElib prototype, is available on GitHub[1].

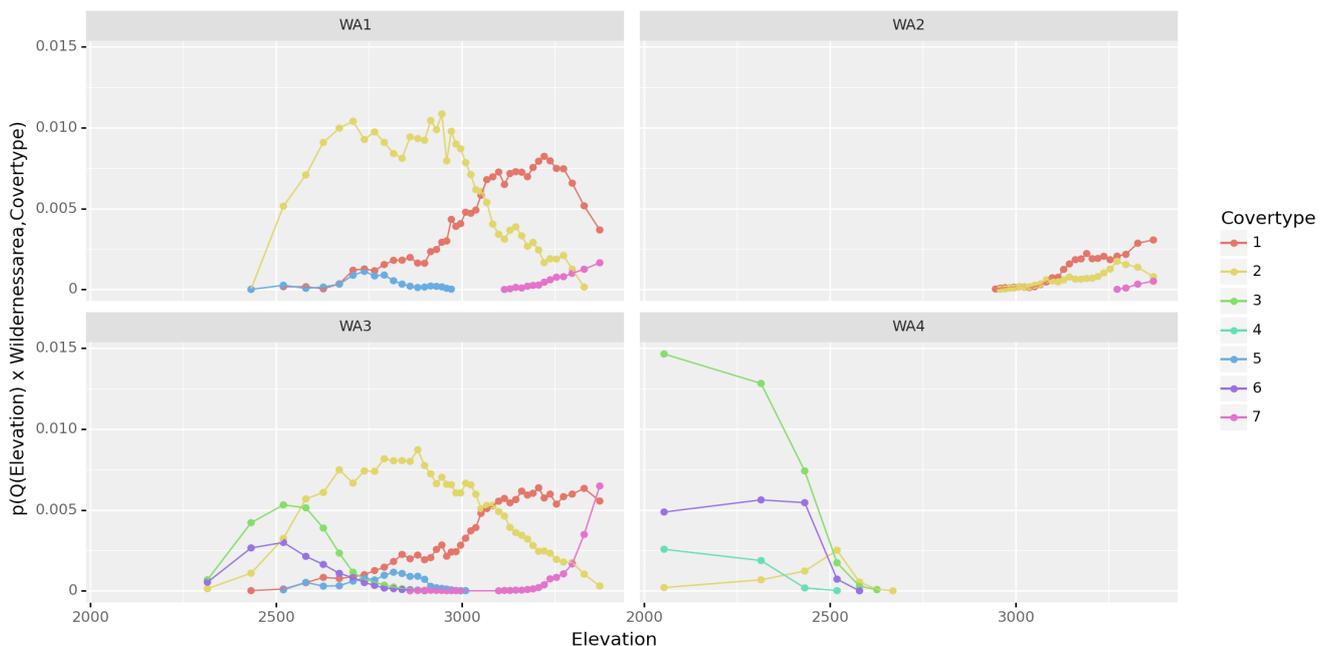

*Figure 2: Output of the SANElib prototype's visual2D function, visualizing a two-dimensional probability estimation of cover types using a combination of the predictive columns Elevation (quantized using Formula (5) with b=39) and Wilderness Area (WA) (reverse pivot of four binary variables in the original dataset). We see that the distribution of covertype over elevation is different depending on the values of WA. For example, cover type 7 is only the most probable class for maximal elevation if wilderness area is WA3. Therefore, we can conclude that a multidimensional combination of values can provide more accurate predictions than naïve Bayesian approaches that falsely assume independence.*

---

[1] https://github.com/SANElibDevTeam/SANElib

## V. QUANTITATIVE EVALUATION

To evaluate our classification method and its implementation, we try to predict forest cover types using the dataset described by Blackard and Dean 1999 [18]. The reason we chose the cover type dataset is it has a reasonably large number of observations, and it contains mixed data (numerical and categorical), both of which makes it a realistic test case. The original data, available on the UC Irvine machine learning repository, encodes categorical values using one binary variable per value (one hot-encoding). However, in our experiment, for the SANElib we reversed the one-hot encoding (reverse pivot) and migrated the data to SQL, resulting in a table with one column per feature. We used a MacBook Pro 2.9 GHz 6-Core i9, 32 GB RAM with MySQL 8, Python 3.9. The benchmarks have been computed using the algorithms of Scikit Learn 0.23.2.

### A. Benchmark

The research described in the original paper[18], using an artificial neural network (ANN), achieved a prediction accuracy of 70.58% on the covertype data, However, the duration of the model was not described. Using this dataset, we computed a benchmark using Gaussian Naïve Bayes (GNB), logistic regression (LR), a decision tree classifier (DTC), and a random forest classifier (RFC). The benchmark algorithms were tested using the original one-hot encoded data. The benchmark for our experiment is summarized in the first five rows of Table 1.

TABLE 1: OBSERVED ACCURACY AND COMPUTATION TIME

| Method | Accuracy | Time |
|---|---|---|
| ANN [18] | .706 | ? |
| Sklearn GNB | .740 | 7.8s |
| Sklearn LR | .823 | 35.1s |
| Sklearn DTC | .936 | 8.0s |
| Sklearn RFC | .946 | 15.6s |
| SANElib MDH EQRB (b=39) | .918 | 21.5s |
| SANElib MDH EWB (b=60) | .927 | 6.9s |

### B. Experiment Procedure and Results

We unpivoted the data using R. Then we loaded the data into a MySQL database server version 8. First, we split the data into training and valuation tables using the MySQL rand function with a seed value of 1 for a train test split of 0.8, meaning that rows assigned random numbers smaller than 0.8 were loaded into the training table, and the others were left out for evaluation. To illustrate the resulting predictive model of our method, using the cover type dataset, a 2d visualization was generated using *elevation and wilderness area* as the feature variable and *cover type* as the target variable on the training table. The result is depicted in Figure 1, generated using the SANElib prototype with the `visual` subroutine. However, to make an actual prediction, we want to consider more dimensions. To evaluate which columns are good predictors, we used the `rank` subroutine of our SANElib prototype, which ranked all columns of the cover type dataset according to Equation (7). The result can be seen in Figure 3. It took 10 seconds to rank all attributes. We see that only the top five variables show more than 0.1 bit of mutual information with the target. Therefore, we calculated our model using these five dimensions, submitting them to the `train` subroutine of our prototype. Thus, a five-dimensional decision table based on a 5-dimensional histogram was generated based on the training data according to Equation (6). For the numerical attributes, we used two different binning methods according to Equation (5a) (EQRB) and (5b) (EWB) in two experiments. We used the same number of bins b for all numeric dimensions for EQRB we used b=39, and for EWB, we used b=60. The values for b were optimized manually. The generated multidimensional probability estimation was then applied to predict the cover type in the evaluation table according to Equation (3) by calling the `predict` subroutine. The result of this computation is shown in . We optimized the number of bins b by trying several values of b, as shown by . We used between 23 and 45 bins in this graphic and plotted the resulting accuracy. The optimum was found at 39 bins. With this configuration, It took 21.5 seconds to complete and showed an accuracy of 91.8%. This result is summarized in the bottom row of Table 1.

```
   OK: Computing mutual information with target

                                         f         mi
0                                Elevation   0.652576
1                                Soil_Type   0.582871
2                          Wilderness_Area   0.271995
3            Horizontal_Distance_To_Roadways 0.129454
4         Horizontal_Distance_To_Fire_Points 0.104075
5                                    Slope   0.051208
6                             Hillshade_9am   0.043945
7           Horizontal_Distance_To_Hydrology 0.032466
8                            Hillshade_Noon   0.030221
9                             Hillshade_3pm   0.029116
10                                  Aspect   0.026773
11             Vertical_Distance_To_Hydrology 0.022152
   Runtime of the program is 11.125906229019165
```

*Figure 3: Output of the SANElib prototype's rank function, resulting in a ranking of features (f) according to mutual information (mi) with the cover_type column (target). The first five features were taken into account for the experiment*

## VI. DISCUSSION

As we can see in Table 1, the accuracy of our proposed method SANElib MDH EWB was 22.1% more accurate than the neural network method reported in the original publication of the Covertype dataset [18]. Our method was 18.7% more accurate than our benchmark using GNB, which means that multidimensional probabilities are far better suited for prediction than assuming independence as NB does. Also, it was 10.4% more accurate than logistic regression. Additionally, our research prototype was only 0.9% less accurate than a decision tree computed by Scikit Learn, a state-of-the-art machine learning toolkit, and only 1.9% less accurate than Scikit's random forest. In terms of computation time, our method MDH EWB was the fastest of all methods. It was 0.9s faster than GNB, 1.1s faster than DTC, 8.7s faster than RFC, 14.6s faster than MDH EQRB and 28.2s faster than LR. The difference between MDH EWB and MD EQRB is that EWB is a parametric quantization method. It relies on empirical extrema, which make the quantization of the evaluation data quite fast. On the other hand, EQRB is a non-parametric method which relies on rank computations.

## VII. CONCLUSIONS

Our experiment with an unpivoted Covertype dataset demonstrated that our proposed IDBML method provides support for mixed data with both numeric and categoric variables. The data in Table 1, from our quantitative benchmark, report that our method was the fastest method, while being almost as accurate as the best algorithms of the current international state of the art. The software prototype and its pilot application in the experiment demonstrate that the principle of Python-based SQL generation for IDBML is feasible, efficient and accurate.

VIII. OUTLOOK AND FUTURE RESEARCH

Because our proposed IDBML method provides support for mixed data, larger-than-memory scalability, and explicability and visualization, and because it is very efficient and almost as accurate as the best algorithms of the current international state of the art in a first experimental quantitative benchmark, we are motivated to do more research in this direction. We took simple approaches to binning (EQRB, EWB), so we conclude there is potential for optimization in better binning methods using SQL, e.g., using 1D clustering or supervised discretization methods. Also, for untrained value combinations we quickly took a majority class predictor, which could be improved by smarter default predictors, e.g., using Bayesian approaches. Further points of future research include experiments on other datasets; tests on a big dataset (e.g., the Steam dataset); implementing more algorithms in Python and SQL using Jinja2 Templates, e.g., linear regression, K-means, and PCA; stabilizing the library using software engineering for industrial capability. Furthermore, we invite the public to contribute. Practitioners could support use cases and pilot applications on their databases, and researchers are invited to contribute implementations to the SANElib In-Database Machine Learning Library.